\documentclass[aip,jap,reprint,superscriptaddress,amsmath]{revtex4-1}
\usepackage{graphicx}
\usepackage{dcolumn}

\begin{document}

%\title{Preparation of freestanding magnetocaloric Heusler thin films and observation of the Martensitic transformation}
\title{Vanadium sacrificial layers as a novel approach for the fabrication of freestanding Heusler Shape Memory Alloys}

\author{L. Helmich}
\email{corresponding author\\e-Mail: lars@physik.uni-bielefeld.de}
\affiliation{Bielefeld University, Department of Physics, Center for Spinelectronic Materials and Devices, 33615 Bielefeld, Germany}
\author{N. Teichert}
\affiliation{Bielefeld University, Department of Physics, Center for Spinelectronic Materials and Devices, 33615 Bielefeld, Germany}
\author{W. Hetaba}
\affiliation{Vienna University of Technology, USTEM, 1040 Vienna, Austria}
\affiliation{Bielefeld University, Department of Physics, Center for Spinelectronic Materials and Devices, 33615 Bielefeld, Germany}
\author{A. Behler}
\affiliation{IFW, Institute for Complex Materials, 01069 Dresden, Germany}
\author{A. Waske}
\affiliation{IFW, Institute for Complex Materials, 01069 Dresden, Germany}
\author{S. Klimova}
\affiliation{Bielefeld University, Department of Physics, Center for Spinelectronic Materials and Devices, 33615 Bielefeld, Germany}
\affiliation{Saratov State University, Saratov, Russia}
\author{A. H\"utten}
\affiliation{Bielefeld University, Department of Physics, Center for Spinelectronic Materials and Devices, 33615 Bielefeld, Germany}

\begin{abstract}
In this study we report a method for the preparation of freestanding magnetocaloric thin films. Non-stoichiometric Heusler alloys Ni-Mn-Sn, Ni-Co-Mn-Sn and Ni-Co-Mn-Al are prepared via sputter deposition. A sacrifial vanadium layer is added between the substrate and the Heusler film. By means of selective wet-chemicals etching the vanadium layer can be removed. Conditions for the crystallization of Vanadium layers and epitaxial growth of the Heusler films are indicated. Magnetic and structural properties of freestanding and as-prepared films are compared in detail. The main focus of this study is on the influence of substrate constraints on the Martensitic transistion.
\end{abstract}

\maketitle

\section{Introduction}
Magnetocaloric materials for room-temperature cooling applications have attracted strong interested mainly for two reasons. On the one hand magnetic cooling devices obviate the need for greenhouse gases as freezing agents. On the other hand there are promising material candidates with high cooling efficiency gains.\cite{Sandeman2012} Magnetocaloric effects (MCE) emerge in any magnetic material due to the interdependence of thermal and magnetic properties. MCE can be induced by application and removal of an external magnentic field.\cite{Planes2009}. One can distinguish between two classes of MCE: Materials which heat up upon magnetization show the direct MCE whereas materials that cool down upon magnetization show the inverse magnetocaloric effect.\cite{Gomez2013} This work is related to three different non-stoichiometric Heusler alloys namely Ni-Mn-Sn, Ni-Co-Mn-Sn and Ni-Co-Mn-Al. All of them are known to show the inverse magnetocaloric effect.\cite{Yuzuak2013, Kainuma2008} Depending on the experimental conditions there are two measures for the MCE. These are the isothermal entropy change $\Delta S$ and the adiabatic temperature change $\Delta T_{\text{ad}}$.\cite{Sandeman2012} Recently there has been a lot a progress in the investigation of $\Delta T_{\text{ad}}$ in bulk materials. Although, to the best of the authors knowledge, this property not been reported yet for Heusler alloys in thin films. However, direct measurements of the adiabatic temperature change in thin films are a challenging task. Thin films with a thickness of hundreds of nanometers provide an amount of mass that is still significantly smaller than the necessary minimum sample sizes even differential scanning calorimeters with high sensitivity would need.\cite{Jeppesen2008} Furthermore thin films are commonly grown on single crystaline substrates with a thickness of hundreds of micrometers. Since these substrates provide a huge heat sink $\Delta T_{\text{ad}}$ cannot be determined directly on these samples. In this study we report on a method in order to elude this issue, i.e. adding a sacrificial Vanadium layer between the substrate and the magnetocaloric Heusler alloy. In a subsequent wet-chemical treatment the vanadium layer can be etched selectively thus resulting in a freestanding Heusler film. 

\section{Experimental details}
Vanadium layers and all Heusler films are grown on MgO(001) single crystalline substrates using a ultrahigh vacuum sputtering system with a base pressure typically better than $1\times10^{-9}$ mbar. The 3 inch sputter sources are arranged in a confocal sputter-up geometry. The distance between target and sample is 21 cm. The Heusler alloy films are deposited from elemental Ni, Co, Mn, Sn, and Al targets. The substrate temperature of $500^\circ\text{C}$ is applied during the deposition process. To ensure a homogeneous stoichiometry the sample is rotated with 5 rpm. Argon flow is regulated to a sputtering pressure of $2.3\times 10^{-3}$ mbar. To prevent surface oxidation all samples are coated with a $2.5$nm MgO capping layer deposited by electron beam evaporation. The stoichiometry of the films is determined by X-Ray fluorescence measurements. Due to the relatively low fluorescence yield of Alumnium the stoichiometry of NiCoMnAl samples are determined by inductively coupled plasma optical emission spectrometry (ICP-OES). In order to obtain depth profiles TEM lamellas are fabricated perpendicular to the sample surface. Subsequently EDX line scans are measured on these lamellas. The crystalline structure is analyzed via temperature dependent X-Ray diffraction. The samples is cooled in a custom-build liquid nitrogen cryostat. Therefore a temperature range from $-150^{\circ}\text{C}$ to $200^\circ\text{C}$ is accessible. XRD is measured in Bragg-Brentano geometry with Cu $K_\alpha$ radiation. Film thickness and density are measured via X-Ray reflectometry. The temperature dependent magnetization is measured using a vibrating sample magnetometer under 10 mT external field applied in in-plane direction. The electrical resistivity is measured using a standard 4-probe setup within a helium-cooled cryostat. The wet-chemical etching procedure is performed with the commercially available acid "`Chromium Etchant No. 1"' by MicroChemicals GmbH. Depending on the preparation conditions of the Vanadium layer it takes five to ten minutes in undiluted acid to remove the Heusler layer completely from the substrate. Afterwards, the brittle layer is washed in deionized water and ethanol. In order to contact the freestanding sample for four-probe measurements a special sample carrier is prepared. This carrier consists of a $\text{SiO}_x$ substrate with four micro-fabricated gold conduction lines on top. The Heusler layer is dried onto this sample carrier. This thin films then sticks to the sample carrier due to wetting effects. This procedure is also convenient to fasten thin films with thicknesses down to $20\text{nm}$ to a TEM grid. This provides practical advantages for TEM imaging since time-consuming preparation steps such as the thinning of the samples thus become unnecessary.

\section{Results and discussion}
Vanadium is an appropriate candidate as a sacrificial layer due to the small lattice mismatch to both the substrate and Heusler alloys. Lattice constants in table \ref{tab:latticeconst} were determined by XRD. It is noteworthy that the lattice constants slightly shift for Vanadium buffered layers, i.e. $5.92\AA$ for NiCoMnSn. 
\begin{table}[h!]
\begin{tabular}{|c|c|}
\hline
Material & lattice const. $\left[\AA\right]$ \\
\hline
Vanadium & $3.05$\\
MgO(100) & $4.21$\\
NiMnSn & 5.98 \\
NiCoMnSn & 5.97 \\
\hline
\end{tabular}
\caption{lattice constants} 
\label{tab:latticeconst}
\end{table}
Vanadium is known to crystallize in a body centered cubic structure and MgO in a face centered cubic structure. Vanadium can be grown epitactically on MgO with a mismatch of $2.3\%$ if the two lattices are twisted by $45$ degree to each other. Also the Heusler alloys can be grown epitactically on Vanadium with a lattice mismatch of $2.0\%$. 

Vanadium deposition at room temperature leads to an amorphous layer. Even post-annealing does not lead to crystallization. A minimum sample temperature of $200^\circ$C is necessary to ensure epitactical growth. An even higher deposition temperature results in a lower surface roughness.  

\begin{figure}[h!]
\includegraphics[width=\columnwidth]{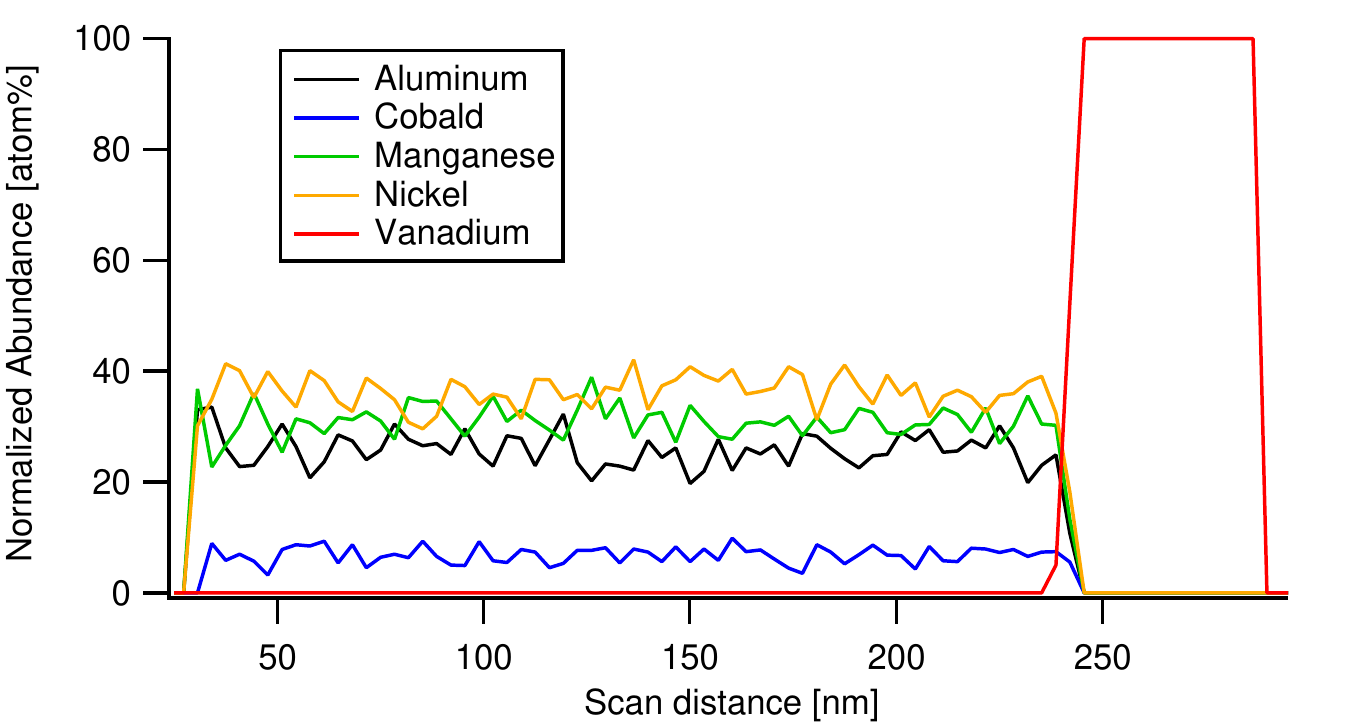}
\caption{EDX linescan on a TEM lamella: Depth profile on NiCoMnAl on V-Buffer. Lines between data points are a guide to the eye.}
\label{fig:edx}
\end{figure}

In former studies of Vanadium as spacer material in a Heusler sandwich structures it is reported that Vanadium is likely to interdiffuse into Heusler materials. This interdiffusion starts above a certain critical temperature and may cause significant problems. \cite{Slomka1999}. In order to determine this critical temperature an ex-situ post annealing study on MgO(001)/V(35nm)/NiMnSn(200nm)/MgO(2.5nm) is carried out. The samples are annealed for one hour each at different temperatures. However, for substrate temperatures higher than $550^\circ$C the Heusler peaks shift to larger lattice constants which indicates undesirable structural changes. Consequently interdiffusion determines an upper temperature limit. A depth profile of a Heusler sample which was deposited at $500^\circ$C substrate temperature is investigated by means of EDX. Results are shown in figure \ref{fig:edx}. Within the range of uncertainties of this method no interdiffusion of Vanadium into the Heusler layer can be observed. This result is in agreement with a depth profile from Sputter-Auger-Electron-Spectroscopy (AES). 

\begin{figure}[h!]
\includegraphics[width=\columnwidth]{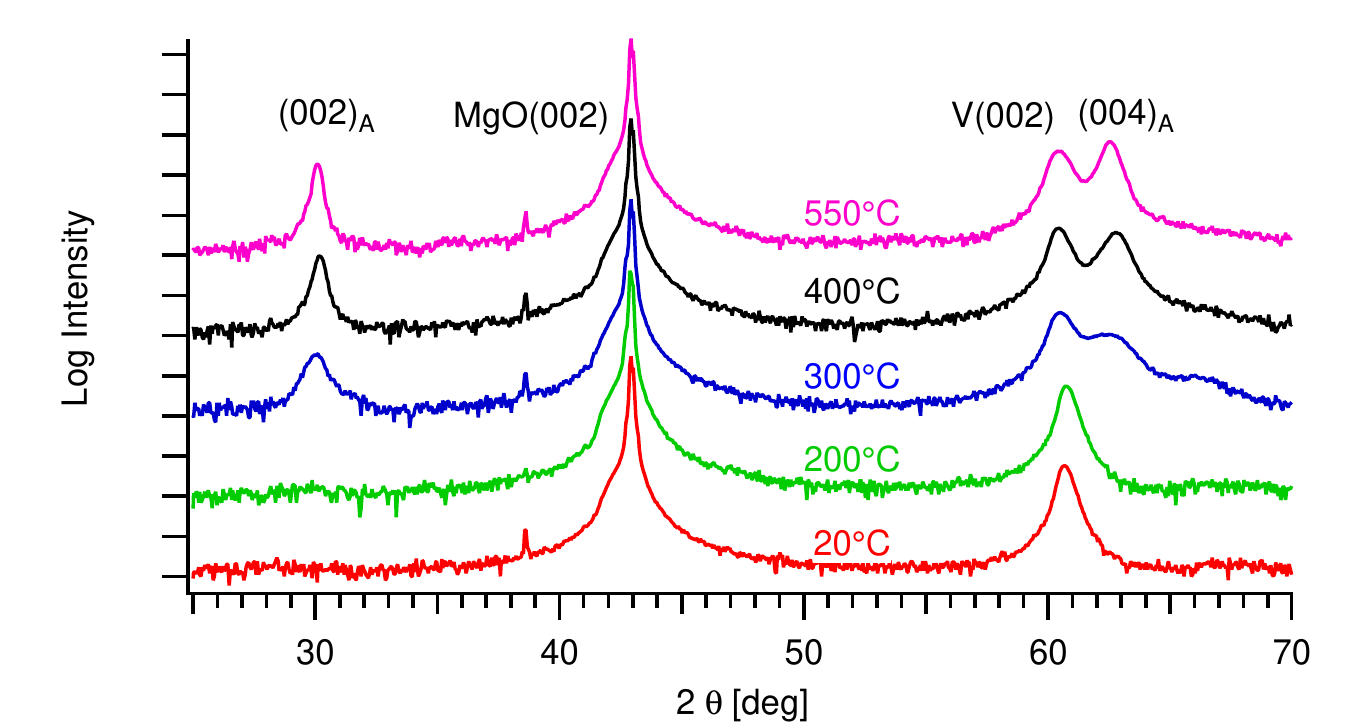}
\caption{XRD pattern of NiMnSn: Temperature series for different substrate temperatures during the deposition of NiMnSn.}
\label{fig:xrd_t_series}
\end{figure}

However, it not possible to fabricate a crystalline Heusler thin film at room temperature. NiMnSn samples are deposited at different substrate temperatures ranging from $20^\circ$C to $550^\circ$C. XRD pattern are shown in figure \ref{fig:xrd_t_series}. The XRD analysis of this temperature series shows that those XRD peaks which belong to the Heusler emerge at substrate temperatures higher than $300^\circ$C.  Therefore the appropriate interval for substrate temperatures during the deposition process is identified between $300^\circ$C and $500^\circ$. Furthermore a depth profile of the samples is measured by means of sputter AES. Within the uncertainties of this method no interdiffusion of Vanadium is observed. 

\begin{figure}[h!]%
\includegraphics[width=\columnwidth]{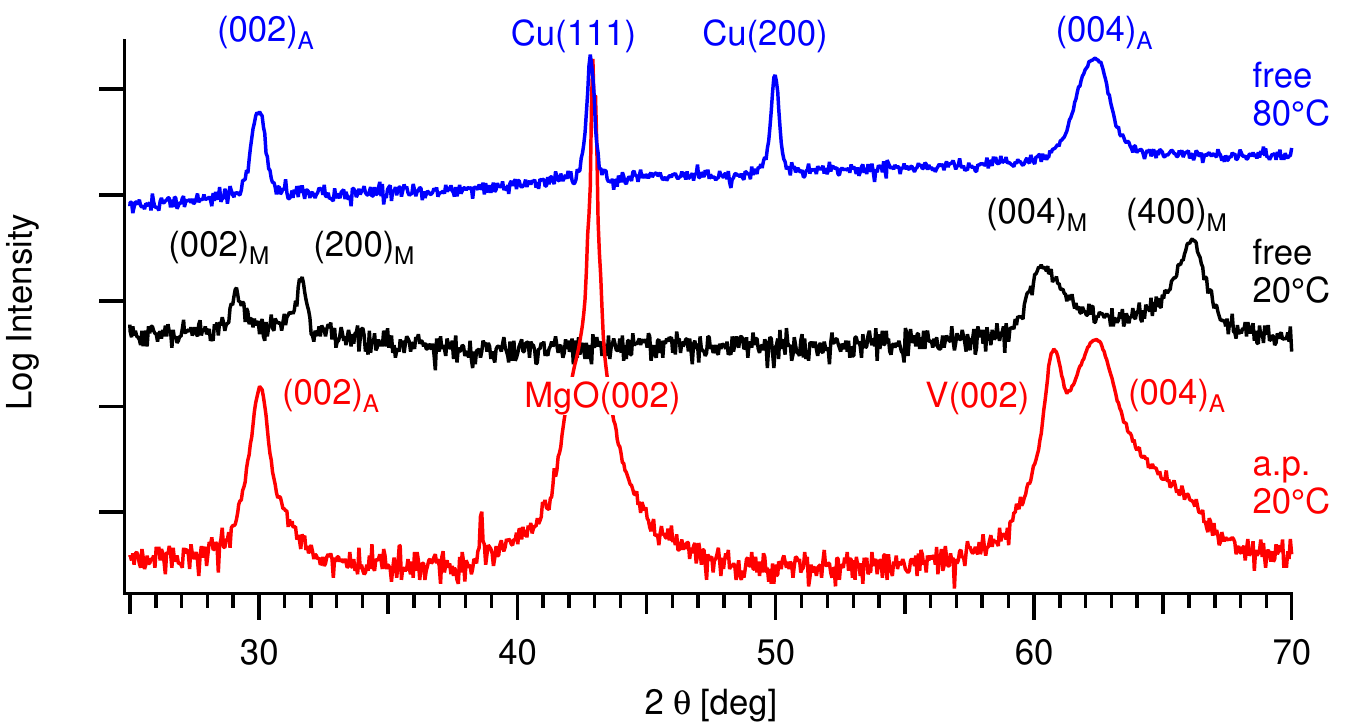}%
\caption{XRD pattern of NiMnSn: as prepared (red), freestanding on glass at $20^\circ$C (black) and $80^\circ$C (blue).}%
\label{fig:xrd1}%
\end{figure}

A comparion of the structural properties between an as-prepared sample and a freestanding sample is carried out on $\text{Ni}_{50.2}\text{Mn}_{34.4}\text{Sn}_{15.4}$. Figure \ref{fig:xrd1} shows the corresponding XRD pattern. The red curve belonging to the as-prepared samples clearly shows a cubic structure, i.e. Austenite, via the $(002)_A$ and $(004)_A$ peaks. The shoulder on the right hand side of the $(004)$ indicates that there is some martensitic contribution. Due to the distortion of the martensitic lattice the $(400)_M$ and $(004)_M$ Martensite peaks can only be observed in a $2\theta-\omega$ scan with a non-zero $\omega$-offset. The black curve is measured on the same sample after removing the substrate and drying the Heusler layer onto a glass sample carrier. The cubic Austenite-Peaks have vanished. Instead four Martensite peaks, namely $(002)_M$, $(200)_M$, $(400)_M$ and $(004)_M$ have emerged. In this case there is no $\omega$-offset needed to measure these peaks since the Heusler film does not dry smoothly on the glass surface but wrinkles. Consequently the film appears similarly to a polycrystalline sample in XRD. Both the red and the black curve are measured at the same temperature, but the crystal structure is clearly different. This means that the Austenite transition temperature is shifted to higher temperatures by removing the substrate. The substrate causes a strain in the Heusler film which hinders the martensitic transition. Hence by removing the substrate it is favorable for the sample to be in the martensitic state. Heating this sample to $80^\circ$ C restores the austenitic structure (blue curve). The two copper peaks result from the copper heating block underneath the sample which was only installed for this measurement. A similar broadening and displacement of the hysteresis due to substrate contraints has already been studied on epitaxial NiMnGa film. \cite{Buschbeck2009}

\begin{figure}[h!]%
\includegraphics[width=\columnwidth]{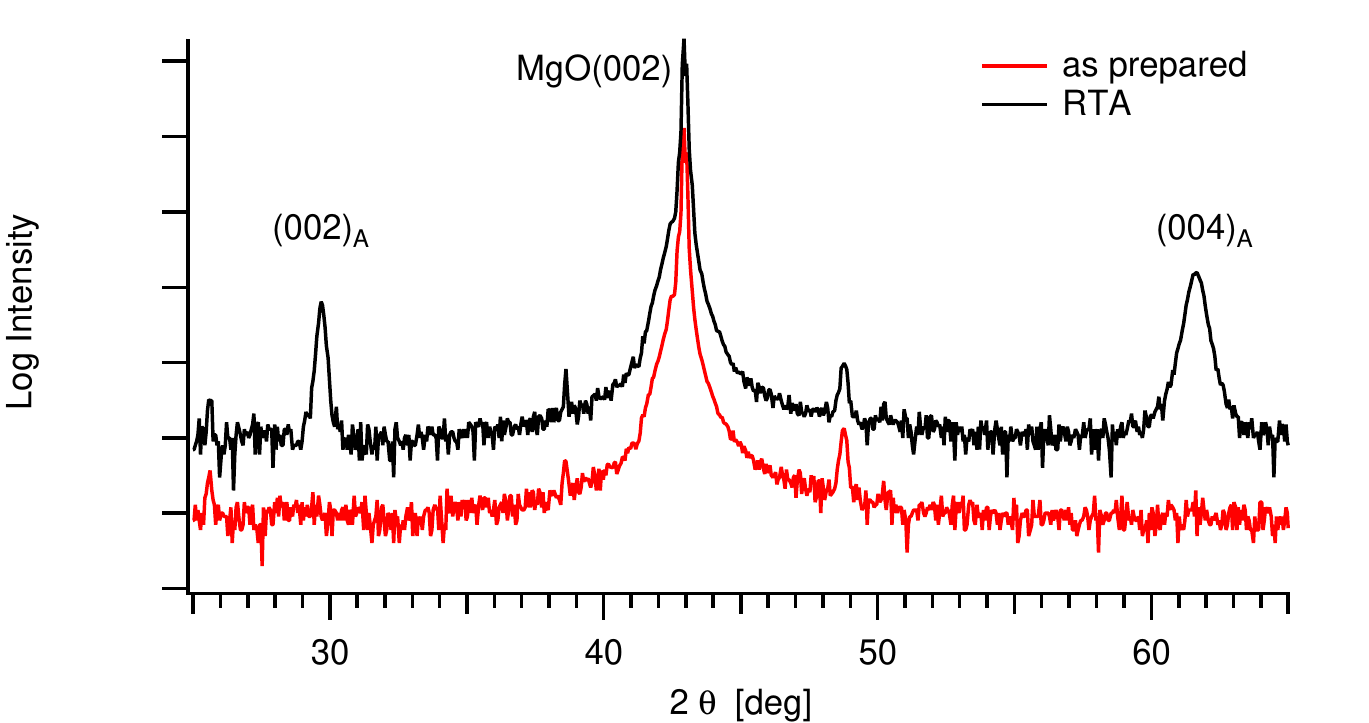}%
\caption{XRD pattern of NiCoMnSn: as prepared (red) and after rapid thermal annealing for 30 seconds (black).}%
\label{fig:rta1}%
\end{figure}

As already mentioned, Vanadium will not crystallize if it is deposited at room temperature. Therefore it is investigated whether a rapid thermal annealing (RTA) process leads to a subsequent crystallization. Figure \ref{fig:rta1} shows the XRD pattern of this investigation on NiCoMnSn. The as-prepared sample (red curve) clearly does not show any crystalline Heusler structure. This sample was exposed to RTA for 30 seconds at 960 Watt resulting in a maximum temperature of $740^\circ$ C. The XRD results of this post-annealing process are shown in black. The distribution of the crystallites is investigated by means of an $\omega$-scan on the cubic (004) peak. This rocking curve shows a FWHM of $1.2^\circ$. The peaks at $2\theta = 48.7^\circ$ are due to $\text{Ni}_3\text{Mn}$-phases. It is noteworthy that there is no Vanadium peak to be seen in the XRD pattern. Thus, RTA is a convenient method to achieve crystallization of NiCoMnSn on an amorphous Vanadium layer. 

\begin{figure}[h!]%
\includegraphics[width=\columnwidth]{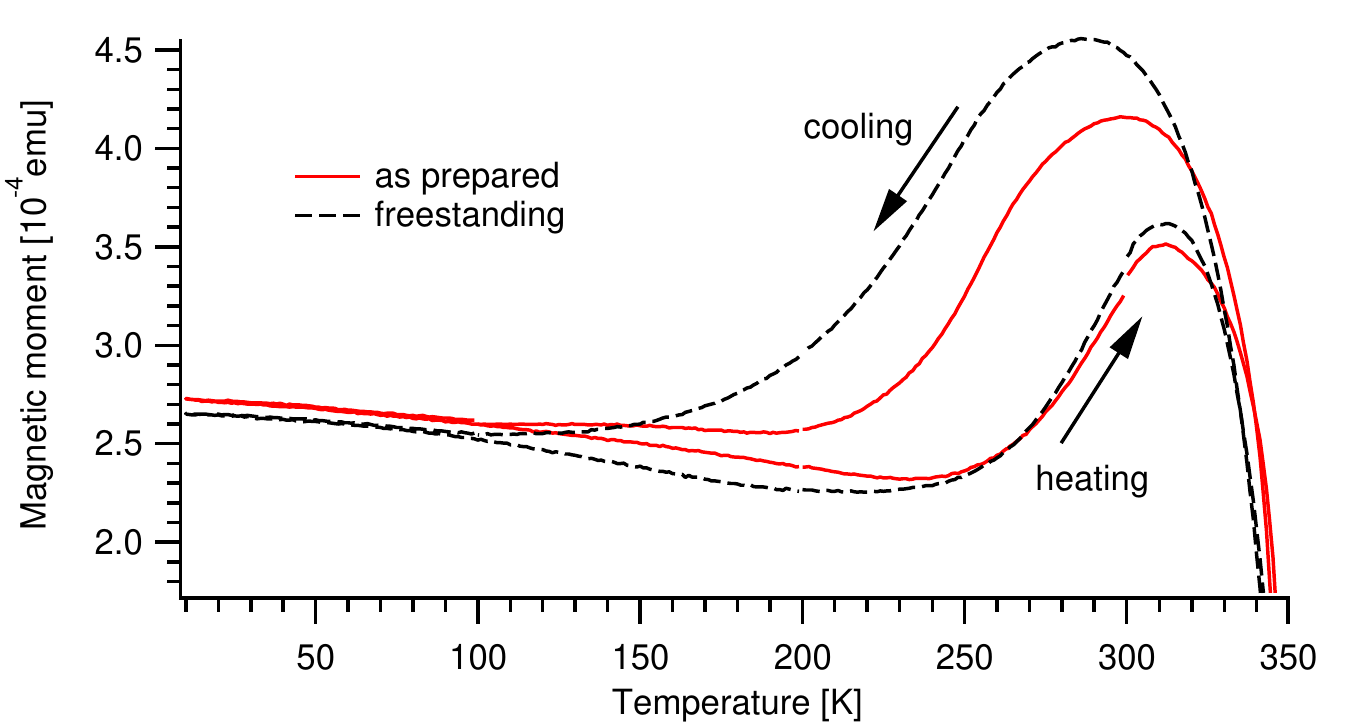}%
\caption{VSM measurement on NiCoMnSn in a constant measurement field of 100 Oe.}%
\label{fig:vsm}%
\end{figure}

The Martensite Phase of NiCoMnSn is known to show a considerably smaller magnetization than its Austenite Phase. Therefore the Martensitic transition can be observed by measuring the magnetic moment of the sample as a function of the temperature. A comparision of the magnetic properties of as-prepard and freestanding NiCoMnSn is carried out by means of vibration sample magnetometry (VSM). Results are shown in figure \ref{fig:vsm}. NiMnSn has a Curie temperature of approximately 300 K\cite{Auge2012}, i.e. room  temperature. Substitution some amount of Nickel by Cobald leads to an increasing Curie temperature. For applications at room temperature a higher Curie temperature is obviously favorable. Both the as-prepared and the freestanding sample show the same behaviour on the heating branch, i.e. Austenite start and Austenite finish temperatures are in good agreement. In contrast the Martensite start temperature is shifted by 10 K and the Martensite finish temperature is shifted by 100 K to lower temperatures resulting in a broadening of the hysteresis. 

\begin{figure}[h!]%
\begin{minipage}{0.59\columnwidth}
\includegraphics[width=\textwidth]{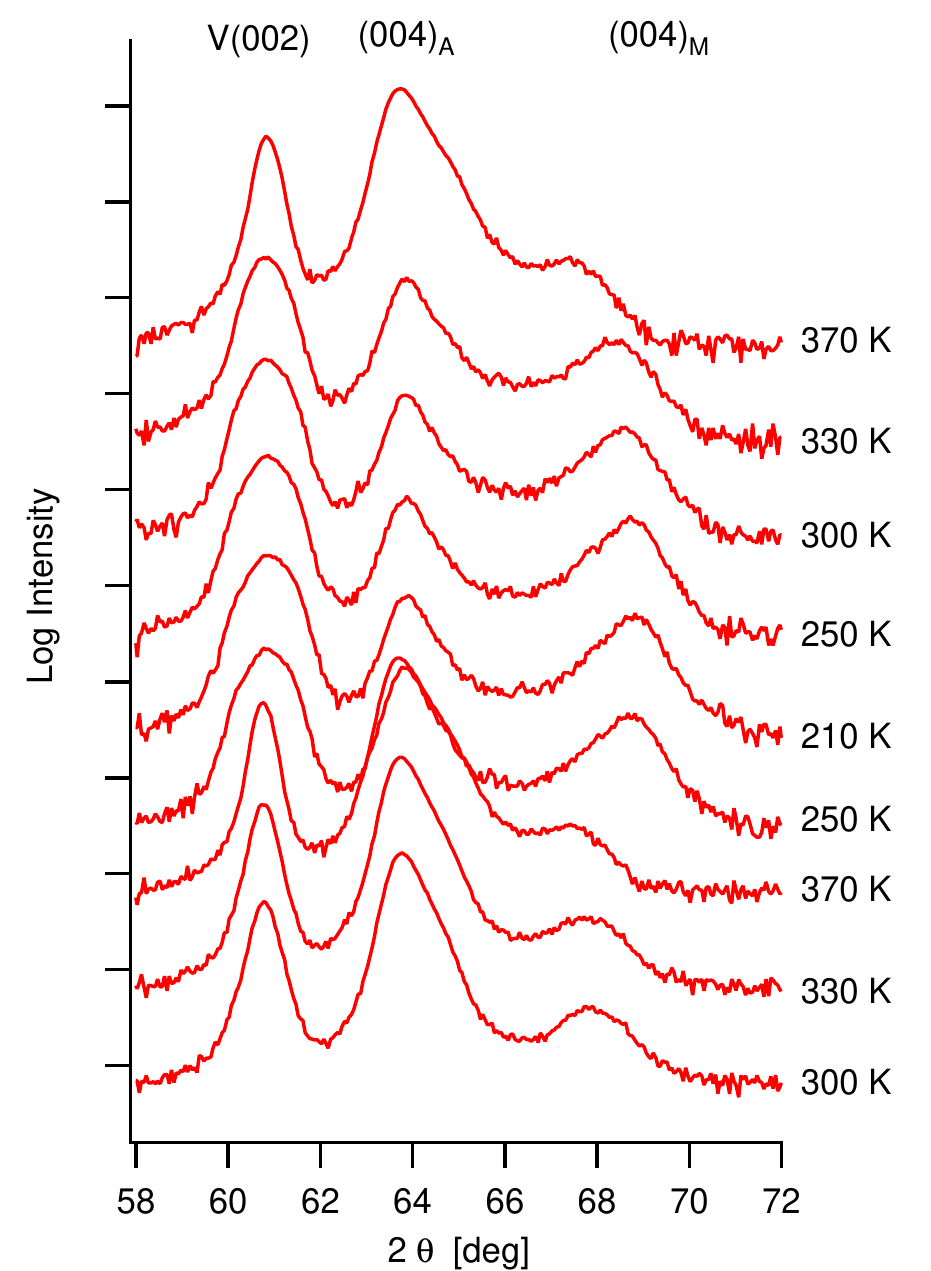}%
\end{minipage}
\begin{minipage}{0.39\columnwidth}
\includegraphics[width=\textwidth]{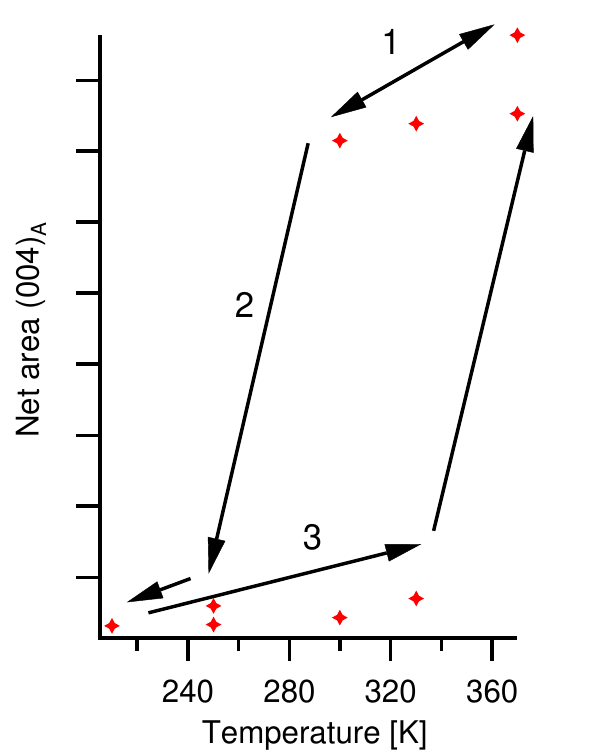}
\end{minipage}
\caption{Temperature dependent XRD of NiCoMnAl. Curves are plotted from bottom to top in the sequence of measurements. (left) The Hysteresis of the Martensitic transformation is extracted form these data. Arrows are only intended as guide to the eye. (right)}%
\label{fig:NiCoMnAl}%
\end{figure}

Magnetic field induced transformation in NiCoMnAl has been studied in bulk materials some time ago \cite{Kainuma2008}. However, we report on the transformation behavior of a thin film $\text{Ni}_{41}\text{Co}_{10.4}\text{Mn}_{34.8}\text{Al}_{13.8}$ sample on a Vanadium buffer which is investigated via temperature dependent XRD. A more detailed study on freestanding NiCoMnAl samples can be found elsewhere\cite{Teichert2014}. 
Results are shown in figure \ref{fig:NiCoMnAl}. The measurement sequence is started at room temperature. At first the sample is heated to 370 K until the net area of the $(004)_A$ peak reaches its maximum value. Subsequently the sample is cooled down to 210 K, i.e. the minimum value of the $(004)_A$ peak and again heated up to $370 K$. The structural hysteresis is obtained from these data by plotting the $(004)_A$ peak net area as a function of the temperature. From the above figure it can clearly be seen that the shape of the Vanadium (002) peaks also changes during the Martensitic transformation. Therefore $\omega$-scans are performed on $2\theta = 60.87^\circ$ for each temperature step. It turns out that the FWHM of these rocking curves also changes from $0.67^\circ$ at 370 K to $0.99^\circ$ at 210 K. This indicates in change in the distribution of crystallites which in turn can be understood a strain effect due to the lattice interaction with the Heusler film. 

\section{Conclusion}  

Freestanding magnetocaloric thin film can be prepared by selective wet-chemical etching of sacrificial Vanadium layers. A multilayer system of Vanadium and a MCE Heusler alloy can be grown epitactically at substrate temperatures between $300\circ$ C and $500^\circ$ C without interdiffusion issues. Releasing the MCE film from the substrate leads to a lowering of the Martensite transition temperature due to the lack of substrate constraints which hinder the transition. Crystalline Heusler film can also be prepared by deposition onto and amorphous Vanadium seed layer and a subsequent rapid thermal annealing process. The differences in the transition behavior can be investigated via VSM. Both Martensite start und Martensite finish temperatures are lower for freestanding films. This broadening of the hysteresis is possibly due to surface oxidation effects. NiCoMnAl magnetocaloric films can also be prepared in thin films. These films excert a strain to the Vanadium buffer during the Martensitic transition.

\section{Acknowledgement}
The authors thank M. Meinert and C. Sterwerf for helpful discussions and K. Fritz for preliminary work on RTA. This work is supported by Deutsche Forschungsgemeinschaft (DFG) in the scope of project A6 within SPP 1599.


\begin{thebibliography}{99}
\bibitem{Sandeman2012} K. G. Sandeman, Scr. Mater. \textbf{67}, 556-571 (2012) 
\bibitem{Planes2009} A. Planes, J. Phys.: Condens. Matter \textbf{67}, 233201 (2009)
\bibitem{Gomez2013} J. R. Gomez, R. F. Garcia, A. M. Catoira, M. R. Gomez, Renew. Sust. Energ. Rev. \textbf{17}, 74-82 (2013)
\bibitem{Yuzuak2013} E. Y\"uz\"uak, I. Diner, Y. Elerman, N. Teichert, A. H\"utten, Appl. Phys. Lett. \textbf{103}, 222403 (2013)
\bibitem{Kainuma2008} R. Kainuma, W. Ito, R. Y. Umetsu, K. Oikawa, K. Ishida, Appl. Phys. Lett. \textbf{93}, 091906 (2008)
\bibitem{Jeppesen2008} S. Jeppesen, S. Linderoth, N. Pryds, L. T. Kuhn and J. B. Jensen, Rev. Sci. Instrum. \textbf{79}, 083901 (2008)
\bibitem{Slomka1999} J.-P. Slomka, M. Tolan, W. Press, M.R. Fritzsimmons, R. Siebrecht, D.W. Schubert, P. Simon, J. Appl. Phys. \textbf{86}, 9 5146 (1999)
\bibitem{Buschbeck2009} J. Buschbeck, R. Niemann, O.Heczko, M. Thomas, L. Schultz, S. F\"ahler, Acta Mater. \textbf{57}, 2516 (2009)
\bibitem{Auge2012} A. Auge, N. Teichert, M. Meinert, G. Reiss, A. H\"utten, E. Y\"uz\"uak, I. Dincer, Y. Elerman, I. Ennen, and P. Schattschneider, Phys. Rev. B \textbf{85}, 214118 (2012).
\bibitem{Teichert2014} N. Teichert et al., arXiv:1410.8583  (2014)
\end{thebibliography}
\end{document}